\pdfoutput=1
\documentclass[prb,twocolumn,showpacs,preprintnumbers,amsmath,amssymb,superscriptaddress]{revtex4-1}

\usepackage{dcolumn}
\usepackage{bm}
\usepackage{amsmath,graphicx}
\usepackage{color}

\newcommand{\tht}{\theta}
\newcommand{\thto}{\tht_{\rm out}}
\newcommand{\thti}{\tht_{\rm inn}}
\usepackage{bm}
\usepackage{url}

\begin{document}
\title{Cross-sectional performance of hollow square prisms with rounded edges}

\author{Hiroyuki Shima}
\email{hshima@yamanashi.ac.jp}
\thanks{(Correspondence author)} 
\affiliation{Department of Environmental Sciences, University of Yamanashi, 4-4-37, Takeda, Kofu, Yamanashi 400-8510, Japan}

\author{Nao Furukawa}
\affiliation{Department of Socio-Environmental Engineering, School of Engineering, Hokkaido University, Sapporo, Hokkaido 060-8628, Japan}

\author{Yuhei Kameyama}
\affiliation{Division of Engineering and Policy for Sustainable Environment, Graduate School of Engineering, Hokkaido University, Sapporo, Hokkaido 060-8628, Japan}

\author{Akio Inoue}
\affiliation{Faculty of Agriculture, Kindai University, Nara 631-8505, Japan}

\author{Motohiro Sato}
\thanks{(Correspondence author)} 
\affiliation{Division of Mechanical and Aerospace Engineering,
Faculty of Engineering, Hokkaido University, Kita 13 Nishi 8, Sapporo 060-8628, Japan}

\date{\today}

\begin{abstract}
Hollow-section columns are one of the mechanically superior structures with high buckling resistance and high bending stiffness. The mechanical properties of the column are strongly influenced by the cross-sectional shape. Therefore, when evaluating the stability of a column against external forces, it is necessary to accurately reproduce the cross-sectional shape. In this study, we propose a mathematical method to describe a polygonal section with rounded edges and vertices. This mathematical model would be quite useful for analyzing the mechanical properties of plants and designing plant-mimicking functional structures, since the cross-sections of the actual plant culms and stems often show rounded polygons.
\end{abstract}


\maketitle

\section{Introduction}

Hollow columns are an excellent functional form as Galileo first reported in the 17th century \cite{Galileo1638}. They exhibit superior mechanical stability and strength against axial compression, bending, and torsion compared to solid columns with the same cross sectional area. This mechanical advantage is mainly due to the improved cross-sectional performance caused by the presence of the internal cavity. In the case of hollow columns, the constituent material is absent in the cavity; instead, it is distributed far from the columnar axis. This material distribution increases the buckling resistance and bending stiffness of the column and reduces the maximum bending stress at the outer edges.
In addition to increased rigidity and strength, hollow columns are also superior to solid columns in that they are lightweight and material-saving. In this way, hollow cylinders and prisms have been used for a long time as building members and machine parts because they have excellent cross-sectional performance and are economical (requires few materials).

Besides artificial structures, the culms, stems and branches of some wild plants have an excellent functional form with hollowness.
Bamboo is a salient example of such the plants \cite{Shima2016,Sato2017}, 
native to warm and moist tropical regions 
in the world \cite{Scurlock2000,InoueFEM2018}.
From a structural mechanics perspective, bamboo culms are considered long hollow cylinders with a slightly thicker base and thinner tips \cite{InoueShimada2019}. This tapered configuration improves rigidity against bending forces caused by cross winds compared to hollow cylinders of uniform diameter.
Furthermore, due to its hollow nature, bamboo culms are lighter than solid columns with the same culm diameter, and therefore can grow faster with a small amount of photosynthates \cite{XSong2016,TMYen2016}. This fast-growing feature allows bamboo to compete favorably for survival with other plants. These facts prove that bamboo is a plant with an excellent functional structure that combines light weight, high stability, and material saving.

\begin{figure}[ttt]
\centering
\includegraphics[width=9.5cm]{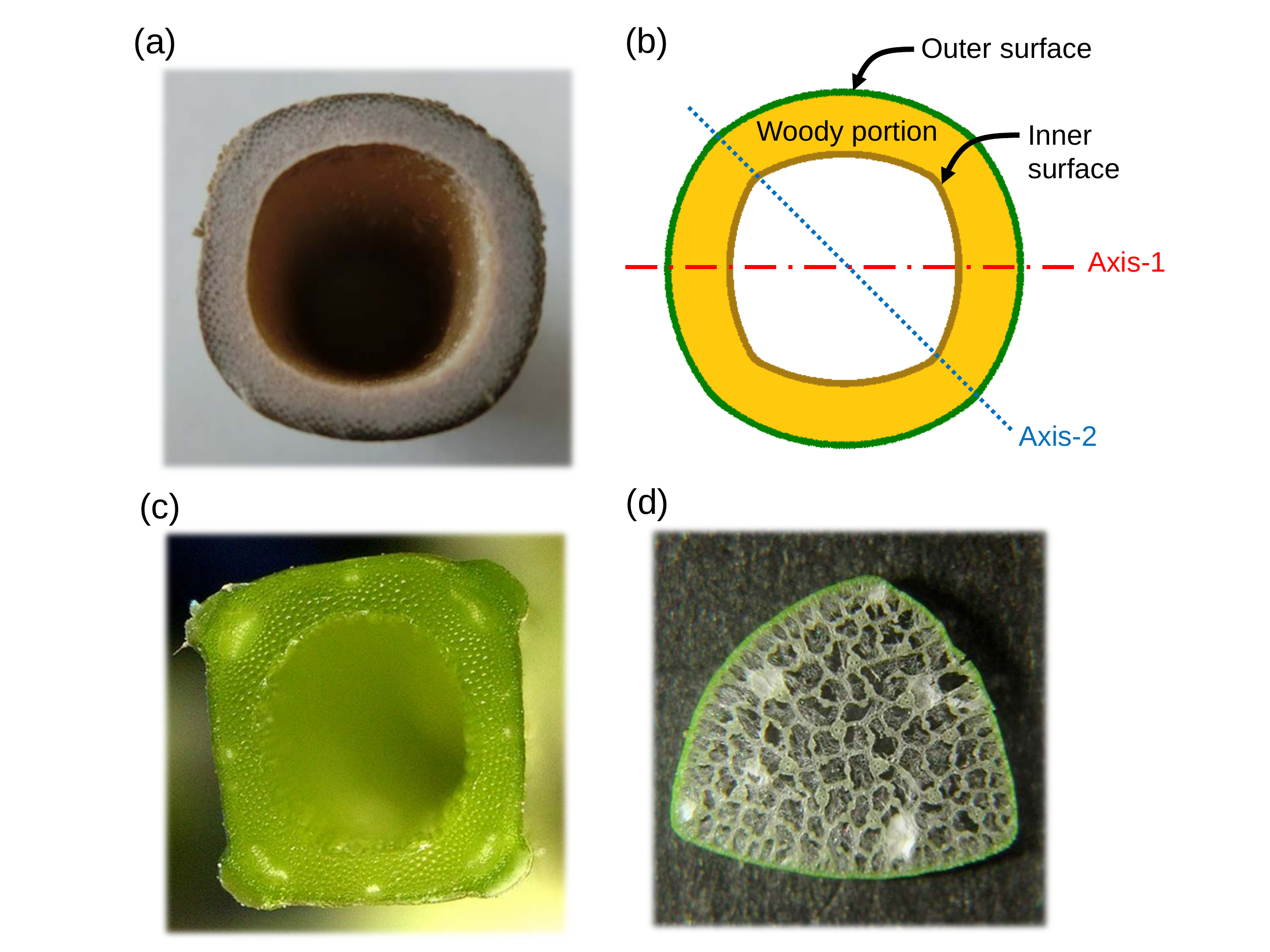}
\caption{(a) Photo of the cross section of a square bamboo: 
{\it Chimonobambusa quadrangularis} (Franceschi) Makino.
(b) Approximate curve of the square bamboo's cross section.
(c) Square cross section of the mint stem; 
{\it Lamium album} var. {\it barbatum}.
(d) Triangular cross section of the papyrus stem; {\it Cyperus microiria}.}
\label{fig01_plant_photo}
\end{figure}

Let us remind that many species of bamboo are endowed with circular cross sections. Nevertheless, certain species of bamboo, like {\it Chimonobambusa quadrangularis} (Franceschi) Makino, have a square-like cross section along the whole length of the culms \cite{Farrelly1984}. Figure \ref{fig01_plant_photo}(a) shows the cross section of the square bamboo \cite{InoueKoshikawa2019}.It is typically two or three centimeters in outer diameter \cite{InoueKoshikawa2019},having a square cross section with rounded sides and filleted corners,
as schematically illustrated by Fig.~\ref{fig01_plant_photo}(b).
Besides the square bamboo, many plants are known to have polygon-shaped cross sections in their stems or branches \cite{GielisBook2017}.
A mint ({\it Lamiaceae}) showing a square stem and a papyrus ({\it Cyperus microiria}) showing a triangular stem are cases in point; 
see Figs.~\ref{fig01_plant_photo}(c)\cite{photo_sq}
and \ref{fig01_plant_photo}(d).\cite{photo_tri}
From the point of view of plant physiology, 
we speculate that the reason why these plants have a polygonal cross-section is that they may promote the formation of phyllotaxis and help identify the location of leaf formation. On the other hand, the choice of polygonal shapes instead of simple circular annulus is expected to give feasible shift in their cross-sectional performance, though no quantitative examination on this issue have been found. 
Moreover, the polygonal cross section of the plant has significantly rounded sides and vertices, unlike the exact regular polygon, which consists of straight line segments and sharp vertices. Therefore, the simple mathematical formulas established in the engineering field cannot be applied to the evaluation of sectional performance. To address this issue, it is essential to develop a theoretical model that describes plant-like polygons with rounded sides and filleted corners.

In the present work, we have created a mathematical model that describes the rounded squares observed in the cross section of a square bamboo. By varying the parameters, this model is broadly applicable to polygonal hollow (and solid) prisms, and can also include triangular, pentagonal and hexagonal prisms. This model was used to analyze the cross-sectional performance of hollow square prisms. Particular emphasis was placed on the section modulus and radius of gyration of the area, which generally determine the mechanical stability and strength of the prism.

\section{Modelling a rounded square with filleted corners}

\begin{figure}[ttt]
\centering
\includegraphics[width=4.5cm]{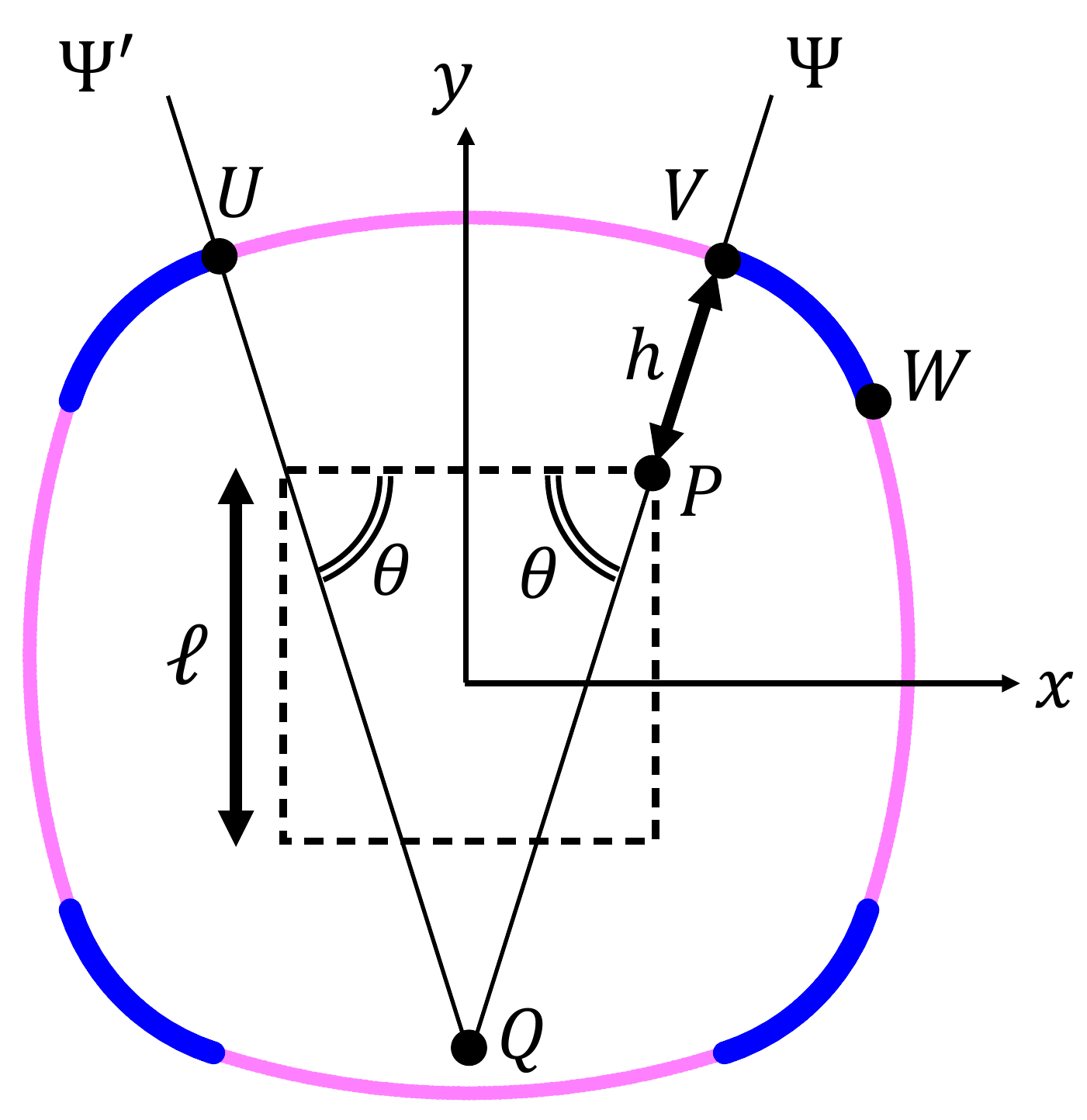}
\caption{A rounded square model with a four-fold symmetry.
Schematic definitions of the three parameters, 
$\ell$, $\tht$, and $h$ are shown.}
\label{fig02_sqmodel}
\end{figure}

We propose a quadrangular prismatic model whose shape is similar that of square bamboo.
Figure \ref{fig02_sqmodel} illustrates the drawing of the cross section,
which possesses the four-fold symmetry composed of four long circular arcs
(colored in magenta) and four short circular arcs (blue).
The cross-sectional shape is uniquely determined by the set of three geometric parameters: $\{\ell, \theta, h\}$, as explained below.
By arranging two of the closed curves with different in a concentric manner,
the hollow cross section as depicted in Fig.~\ref{fig01_plant_photo}(b) can be obtained.

To obtain the model, we first prepare a reference square with the side length $\ell$, 
which is depicted by a dotted square in Fig.~\ref{fig02_sqmodel}.
Next, we draw a straight line $\Psi$ that passes through the upper-right vertex, P, 
of the reference square
such that it forms an angle $\theta$ with the horizontal side of the square.
Another straight line $\Psi'$ is also drawn at the symmetric position
to the line $\Psi$ with respect to the $y$-axis.
The intersection of the two lines is denoted by Q.
Then, we draw a circular arc around Q, which is represented by UV in Fig.~\ref{fig02_sqmodel}.

The radius of the circular sector UVQ is defined by the sum of the line segment length PQ
and $h$, as indicated in the figure.
Another circular arc around P with radius $h$, represented by VW in Fig.~\ref{fig02_sqmodel},
is also drawn.
Finally, we perform the similar drawing of circular arcs as above
for other three vertices of the reference square
to create the rounded square cross section with filleted corners.

The variable range of the angle $\tht$ is defined to be $\pi/4 \le \tht \le \pi/2$.
Particularly when $\tht=\pi/4$, the small arc VW converges to a point so that
the rounded square becomes a true circle with a radius of $(\ell/\sqrt{2})+h$.
On the other hand, when $\tht=\pi/2$, the large arc UV be 
a horizontally straight line segment and VW becomes a quadrant arch.

\begin{figure*}[ttt]
\centering
\includegraphics[width=\textwidth]{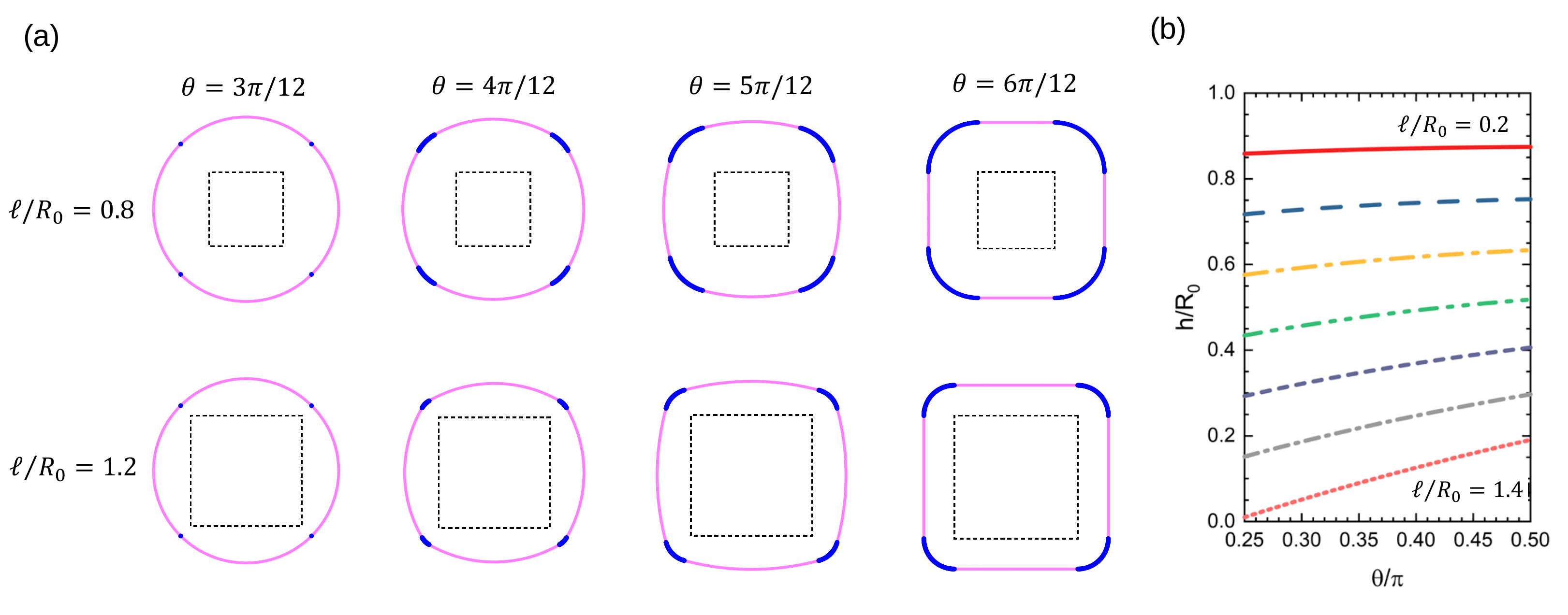}
\caption{(a) Geometric variation of a rounded square
due to changes in variables $\tht$ and $\ell$
under the condition where the enclosed area $A=\pi a_0^2$ is fixed.
The constant $a_0$ serves as the unit of length.
Dotted lines show the reference squares with side length $\ell$.
(b) Dependence of the small sector's radius $h$
on $\tht$. The value of $\ell/a_0$ is varied with interval of $0.2$.}
\label{fig03_variation}
\end{figure*}

The total area enclosed by the rounded square, designated by $A$,
is a function of the three parameters $\ell$, $\tht$, $h$.
Unless $\tht = \pi/2$, it is explicitly written by 
\begin{eqnarray}
A(\ell,\tht,h) 
&=& (2\pi-4\tht) \left( \frac{\ell}{2 \cos \tht} + h \right)^2 
+ ( 1-\tan \tht )\ell^2 \nonumber \\
& &+ (4\tht-\pi) h^2.
\label{eq_001}
\end{eqnarray}
Only when $\tht=\pi/2$, Eq.~(\ref{eq_001}) does not work as $(\cos\tht)^{-1}$
and $\tan\tht$ diverge; instead it is replaced by
\begin{equation}
A\left( \ell, \tht=\frac{\pi}{2}, h \right) = \pi h^2 + 4\ell h + \ell^2.
\label{eq_001dash}
\end{equation}

Note that the model possesses the tangent continuity at any point on the closed curve.
In addition, the model is essentially different from super-ellipses \cite{LameBook1818,PShi2015},
being expressed in terms of $p$-norm and the vector notation 
by $\|\bm{r}-\bm{r}_0\|_p = {\rm const}$,
or its generalization called a Gielis curve \cite{Gielis2003,Gielis2012,PShi2020},
while both of them are known as powerful tools for describing the natural shape of plants.
Furtheremore, a method of drawing a rounded triangle 
by using a complex function has been proposed quite recently \cite{Jafari2020}, 
but again our method is essentially different from it.

A notable feature of the modeling we have presented is based on a patchwork-like method, where four pairs of long and short arcs are sequentially jointed to form a closed curve of the rounded square. Due to this feature, it is possible to introduce asymmetry in the closed curve, as will be discussed later. In other words, we can change the curvature of the sides of the polygon and the degree of rounding of the vertices at each location. In addition, the constituent arcs are represented by simple functions, making it easy to calculate exact solutions for cross-sectional performance. These features are expected to be quite useful when analyzing the mechanical properties of polygonal culms and stems that actually exist in nature.

It should be also emphasized that
attempts to mathematically reproduce the complex morphology of plants have 
great academic significance \cite{Klingenberg2015,Schmidt2018,PShi2018,LCao2019,Gielis2020growth}
and high applied value from the perspective of biomimetic technology \cite{Dargahi2019,Fiorello2020}.
In this respect,
the mathematical model proposed in this paper can be expected to contribute to obtain
better understandings of the mechanical properties of plants with polygonal culms and stems
and to develop plant-mimetic optimal design of high-rise buildings and hollow pipe structures.

Figure \ref{fig03_variation}(a) shows how the geometry varies with changing the values
of $\ell$ and $\theta$.
In the drawing, we fixed the area to be constant as $A=\pi a_0^2$,
and the constant $a_0$ was used as the unit of length in the model.
Under such the constant-area condition,
$h$ is not an independent variable but a function of $\ell$ and $\tht$,
as is followed from Eqs.~(\ref{eq_001}) and (\ref{eq_001dash}).
Figure \ref{fig03_variation}(b) shows the dependence of $h$ on $\tht$
for various values of $\ell$,
indicating that $h$ increases monotonically with $\tht$
for every $\ell$.
Particularly when $\ell=\sqrt{2} a_0$,
the curve passes through the point $(h,\tht)=(0,\pi/4)$.
For larger $\ell$, the $h$-curve has an intersection with the $\tht$-axis
at $\tht=\tht^* (> \pi/4)$.
As a result, the definition range of $\tht$ becomes limited to $\tht^* \le \tht \le \pi/2$
with the lower limit $\tht^*$
that is an increasing function of $\ell$.
Eventually when $\ell=\sqrt{\pi}a_0$,
$\tht^*$ reaches $\pi/2$ so that
the curve shrinks to a point located at $(h,\tht)=(0,\pi/2)$
that represents a strict quadrangle with four right-angle corners.

Two square closed curves with rounded sides and filleted corners,
drawn by the method abovementioned,
are arranged in a double concentric manner
to create an approximate curve that resembles the cross section of hollow square prisms
found in square bamboos and other plants.
An example was shown in Fig.~\ref{fig01_plant_photo}(b).
The outer and inner closed curves that enclose the woody portion of the square bamboo are 
constructed by the parameter settings:
$\ell=1.0$, $\tht=6\pi/20$ and $h=0.6$ for the outer curve,
and
$\ell=1.0$, $\tht=7\pi/20$ and $h=0.2$ for the inner curve,
respectively,
with a common unit of length $a_0$.
In the case of actual square bamboo, 
the shape of the cross section changes 
depending on the height from the ground to the tip.
We have confirmed that many of them can be reproduced 
by setting appropriate parameter values $\{\ell, \tht, h\}$.

\section{Cross-sectional performance}

Our current aim is to evaluate the performance of rounded square cross sections
observed in square bamboo culm and other plant stems.
Generally for a given column,
the cross-section performance measures the degree of contribution
from the cross-sectional shape to the mechanical stability and strength
of the column against external forces.
It is characterized by four 
morphological quantities evaluated from the cross-sectional shape:
the area $A$, the second moment of area $I$,
the gyration radius of area $R_g=\sqrt{I/A}$, and the section modulus $Z=I/e$.
Here, $e$ is the greatest distance from an axis assigned to the cross section
to the extreme edge of the outer enclosing curve.

The following discussion will focus on how the latter two quantities,
$R_g$ and $Z$,
depend on the geometry of the cross section,
because the two are directly relevant to the mechanical stability and strength
of the column, respectively.
In plain words,
$R_g$ measures the buckling resistance of a column under axial compression,
and $Z$ measures the yielding strength of it;
these points will be revisited later.
The actual calculation method of $R_g$ and $Z$ will be explained in Appendices A and B.
It should be noted that in general,
the values of $R_g$ and $Z$
are dependent on the definition of the axis assigned to the cross section.
In the present study,
we pay attention to the two different axis configurations
as indicated in Fig.~\ref{fig01_plant_photo}(b),
with respect to which the cross sectional performance is examined.
We shall find that $I$ (and $R_g$) take the identical value
for the two axes, 
even when the axis is rotated by an arbitrary angle in the same plane,
because of the four-fold symmetry of the cross section to be considered;
see Appendix C for details.

The effect of geometric variation on $R_g$ can be addressed
by introducing the improvement ratio, $\eta$, defined by 
\begin{equation}
\eta = \frac{R_{gw} - R_{gw}^0}{R_{gw}^0}.
\end{equation}
Here, $R_{gw}$ is the gyration radius of the woody portion 
sandwiched by two concentrically arranged rounded squares;
see Fig.~\ref{fig01_plant_photo}(b).
$R_{gw}^0$ is the gyration radius of an annulus, obtained by $\tht_{\rm inn}=\tht_{\rm out} = \pi/4$,
having the same area as that of the rounded square-shaped woody portion.
The improvement ratio $\eta$ measures how $R_{gw}$ increases compared with 
the case of a simple annulus.
When $\eta$ takes a large absolute value with positive (or negative) sign,
it means that the gyration radius has increased (decreased) 
due to the change in cross-section shape from the simple annulus to a rounded square.

Similarly, we also define the improvement ratio of the section modulus 
of the system, $\zeta$, defined by 
\begin{equation}
\zeta = \frac{Z_w - Z_w^0}{Z_w^0}.
\end{equation}
Similar to the case of gyration radius,
$Z_w$ is the section modulus of the woody portion,
and $Z_w^0$ is that of an annulus having the same area as the woody portion.
Considering the dependence of $Z_w$ on the direction of the axis,
as well as taking into account the four-fold symmetry of the cross sectional shape,
we chose the specific two axis configurations, labeled by the axis-1 and axis-2,
which are illustrated in Fig.~\ref{fig01_plant_photo}(b).

\section{Numerical conditions}

Each of the two concentric rounded squares
has a set of four geometric parameters:
$\{\ell_{\rm out}, \tht_{\rm out}, h_{\rm out}, a_{\rm out}\}$
for the outer square, and 
$\{\ell_{\rm inn}, \tht_{\rm inn}, h_{\rm inn}, a_{\rm inn}\}$
for the inner square.
For both squares, 
the areas $A_{\rm out} = \pi a_{\rm out}^2$ and 
$A_{\rm inn} = \pi a_{\rm inn}^2$ are set to be unchanged
in order for the area between the two concentric squares, $A_{\rm out}-A_{\rm inn}$,
to remain constant, too.
These constant-area conditions make each of $h_{\rm out}$ and $h_{\rm inn}$
be dependent on the remaining three parameters, respectively.

In the following discussion,
the relative ratio of $a_{\rm out}$ to $a_{\rm inn}$
is set to $a_{\rm out}/a_{\rm inn} = 1.5$, as a case study,
without loss of generality of our conclusion,
while $a_{\rm inn}$ will be used as a unit of length.

\section{Results: Improvement ratio}

\begin{figure}[ttt]
\centering
\includegraphics[width=8.5cm]{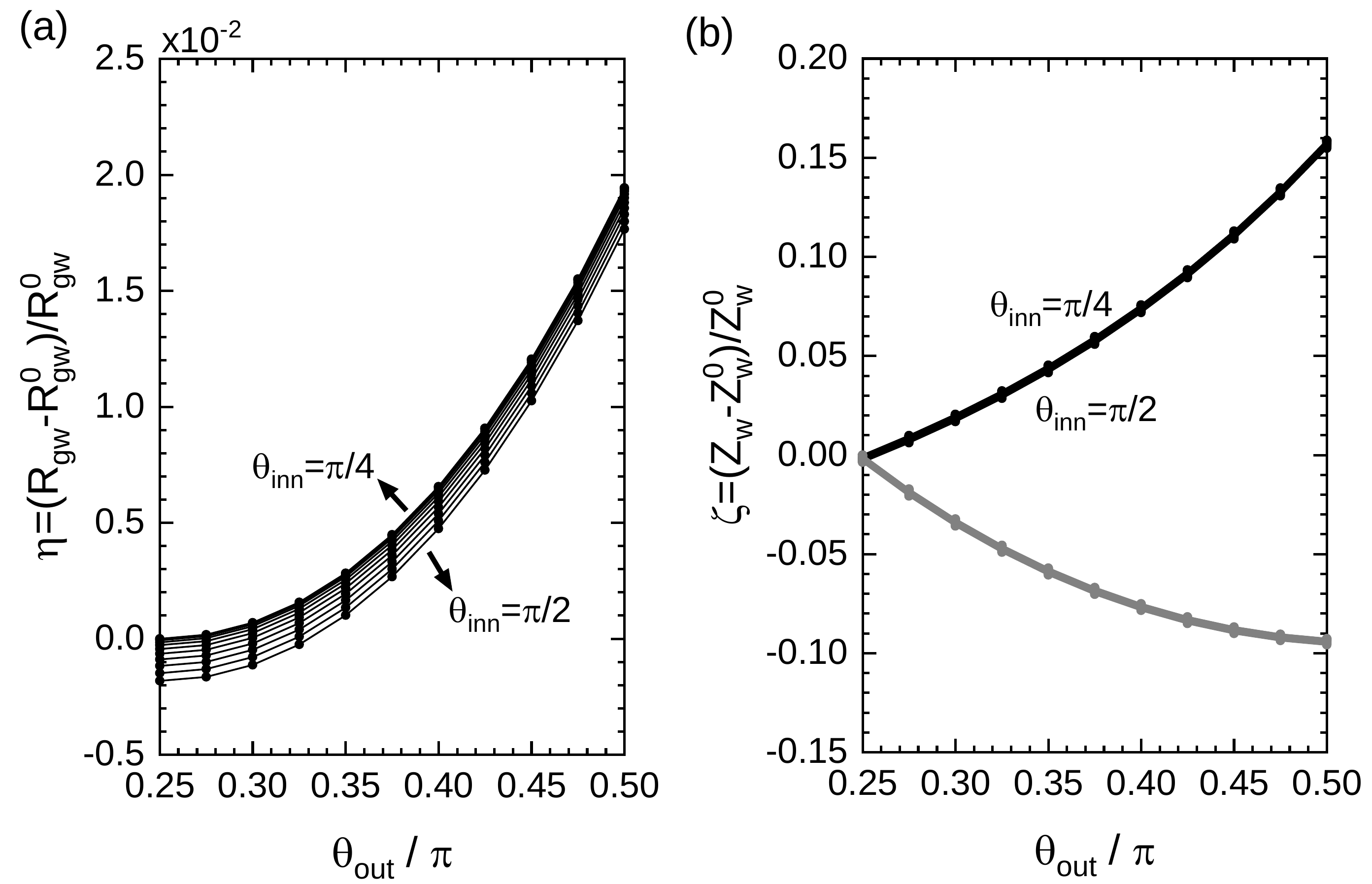}
\caption{Improvement ratios, $\eta$ and $\zeta$, for 
the case of $\ell_{\rm out}/a_{\rm out}=1.2$ and $\ell_{\rm inn}/a_{\rm inn}=0.8$.
(a) $\eta$ as a function of $\tht_{\rm out}$.
(b) $\zeta$ as a function of $\tht_{\rm out}$.
Upper (black) and lower (gray) branches 
correspond to the results with respect to
the axis-1 and axis-2 depicted in Fig.~\ref{fig01_plant_photo}(b), respectively.}
\label{fig04_GyR_out12}
\end{figure}

To examine the cross-sectional performance of the system,
we have changed the values of $\ell_{\rm out}$, $\ell_{\rm inn}$, $\tht_{\rm out}$, and $\tht_{\rm inn}$
in a systematic manner to obtain the geometric dependences of $\eta$ and $\zeta$,
only a part of which will be shown below.

Figure \ref{fig04_GyR_out12}(a) shows 
the improvement ratio $\eta$ as a function of $\thto$ and $\thti$.
Other geometric parameters are set to be
$\ell_{\rm out}= 1.2$ and $\ell_{\rm inn}=0.8$ in unit of $a_{\rm inn}$.
It is observed that
$\eta$ grows with $\thto$ in a nearly parabolic manner
without local maximum peak
for every choices of $\thti$,
and the effect of $\thti$ variation is insignificant
under the present condition.
The maximum value of $\eta$ can be obtained
at $\thto=\pi/2$ and $\thti=\pi/4$,
which corresponds to the hollow cross section composed of
a quadrangular-shaped outer curve
and a circular-shaped inner curve.
We see that the maximum increment in $\eta$ is a few percent at most,
which will be intuitively understood by an analytic estimation discussed later.

Figure \ref{fig04_GyR_out12}(b) shows the improvement ratio $\zeta$
for the section modulus.
The upper branches (almost all of the curves with different $\thti$ appear to be superimposed)
correspond to the results with respect to the axis-1,
and the lower branches do the axis-2.
In the upper branches,
a similar trend as that of $\eta$ is observed,
while the insensitivity to the change in $\thti$ is more prominent.
The increment obtained at the maximum is estimated to be 16 percent,
which will give feasible contribution to the mechanics of the hollow columns.
This increment is attributed to the elongation of the distance $e$
from the axis to the extreme edge of the outer boundary,
which is a direct consequence of the geometric variation in the cross section
from the simple circle to a rounded square.

\begin{figure}[ttt]
\centering
\includegraphics[width=8.5cm]{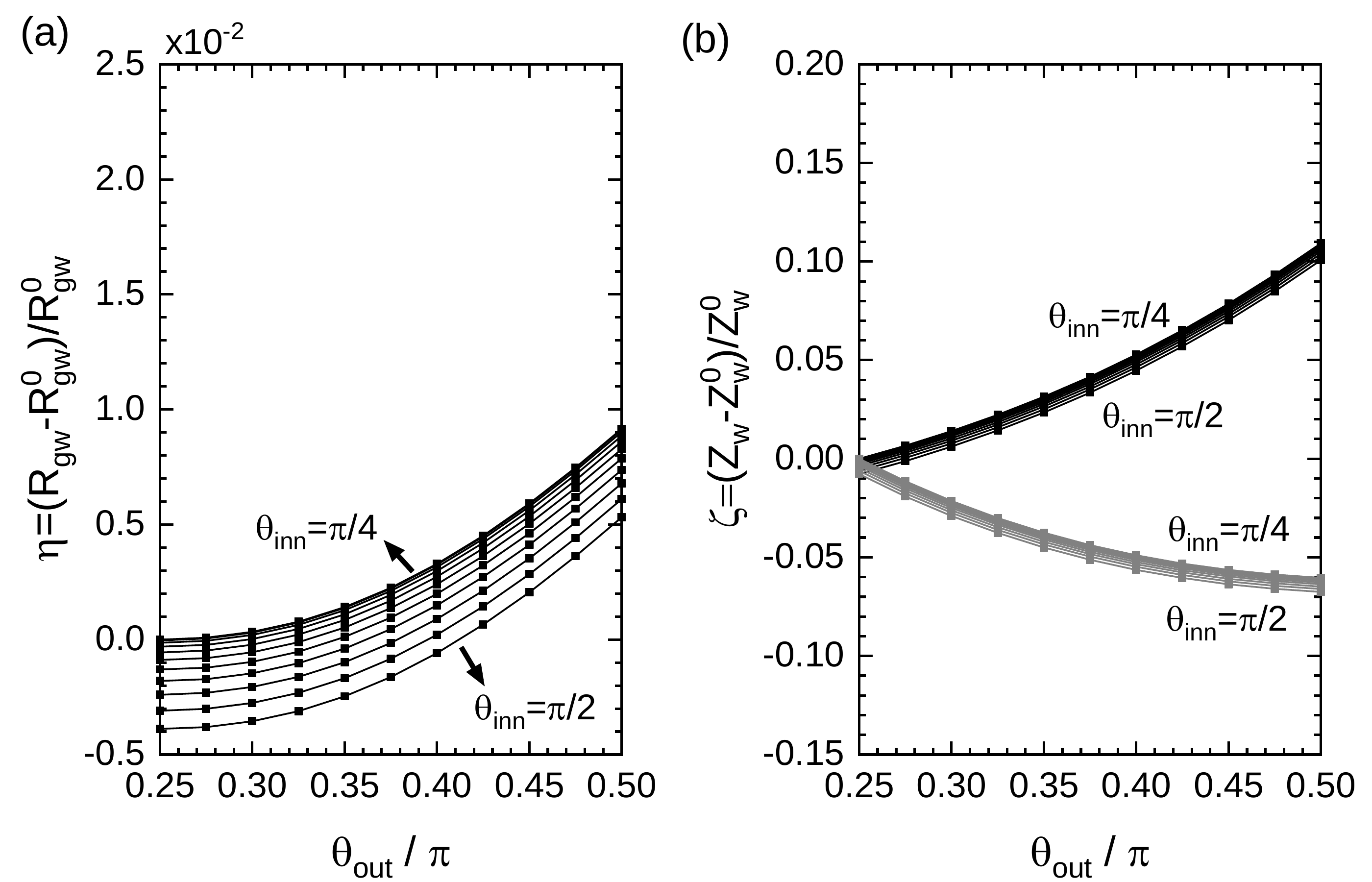}
\caption{Improvement ratios for 
the case of $\ell_{\rm out}/a_{\rm out}=0.8$ and $\ell_{\rm inn}/a_{\rm inn}=1.2$.
(a) $\eta$; (b) $\zeta$.}
\label{fig05_GyR_out08}
\end{figure}

If the magnitude relation between $\ell_{\rm out}$ and $\ell_{\rm inn}$ is altered,
the maximum increments in $\eta$ and $\zeta$ are both depressed.
An example is shown in Fig.~\ref{fig05_GyR_out08},
in which $\ell_{\rm out}=0.8$ and $\ell_{\rm inn}=1.2$ are set
in unit of $a_{\rm inn}$.
It thus turns out that,
for obtaining a greater $\eta$ and $\zeta$,
it is advantageous to make the outer curve more square while making the inner curve more circular.

\section{Discussion}

\subsection{Variable range estimation of $\eta$ and $\zeta$}

Given the ratio of $a_{\rm out}/a_{\rm inn}$,
the maximum and minimum values of $\eta$ can be roughly estimated
by considering the following two extreme situations:
i) the outer curve is an exact quadrangular square with the enclosed area of $\pi a_{\rm out}^2$
and the inner curve is a simple circle with the enclosed area of $\pi a_{\rm inn}^2$,
and 
ii) vice versa ({\it i.e.,} the outer circular curve and the inner quadrangular  curve
with the definitions of the areas same as i)).
Suppose that $a_{\rm out} = p a_{\rm inn}$ with a proportional constant $p$.
We then have
\begin{equation}
\eta = \sqrt{\frac{(\pi/3)p^4 - 1}{p^4 -1}}-1 \;\; \mbox{for case i)},
\end{equation}
and
\begin{equation}
\eta = \sqrt{\frac{p^4 - (\pi/3)}{p^4 -1}}-1 \;\; \mbox{for case ii)}.
\end{equation}
Substituting $p=1.5$ in accord with our numerical condition in Sec.IV, 
we obtain $\eta\simeq 0.03$ for the case i)
and $\eta\simeq -0.006$ for the case ii).
Therefore, $\eta$ should take an intermediate value between the two extrema,
as is consistent with our numerical results shown in 
Figs.~\ref{fig04_GyR_out12}(a) and \ref{fig05_GyR_out08}(a).
A similar argument holds true for the maximum and minimum values of $\zeta$.

\subsection{Effect of fillet at vertices}

It has been proved that
the cross-sectional performance is highest 
when the outer boundary is a quadrangle with sharpened vertices
and the inner boundary is a circle \cite{Gere1972}.
In practice, however, a strict quadrangle is not preferred as the outer boundary.
This is because when the hollow prism is subjected to bending deformation, 
stress concentrations occur around acute vertices so that it locally breaks.
To prevent the local breaking, it is better to fillet corners as suggested in the present model.
We have demonstrated that such the filleted corners do not significantly reduce the cross sectional performance of the hollow square prism, thus achieving the trade-off between the ideal performance obtained by sharpening the corners at the outer boundary and the suppression of the fragility at the corners realized by rounding them. This may be the wisdom of wild plants, while a certain physiological reason should also be possible for the preference of filleted corners.

It also should be noted that, even if the corners are filleted to some extent, the bending stress may be concentrated at the corner, causing premature failure. In order to quantitatively clarify the stress distribution near the corners, it is necessary to analyze the effect of stress triaxiality or perform finite element modeling, which is an interesting problem left for the future.

\subsection{Implication for the mechanical rigidity and strength}

The relationship between cross-sectional shape and column mechanical stability and strength 
may need to be explained in more detail for readers outside the field.
In structural mechanics, 
the gyration radius of area, $R_g$, is known as a key quantity for comparing
the buckling resistance of elastic columns with different cross-sectional shapes.
Long columns often buckle when it receives an axial stress
that exceeds a certain threshold value.
The maximum compressive stress the column can withstand is called 
a buckling stress.
An important fact is that
the magnitude of buckling stress is determined by $R_g$;
more concretely, the larger $R_g$, the greater the buckling stress.
The Euler column formula can be used to analyze for buckling of 
a long column with a load applied along the central axis:
\begin{equation}
\sigma_{\rm cr} = \frac{\pi^2 E}{c (L/R_g)^2}.
\label{eq_euler}
\end{equation}
In Eq.~(\ref{eq_euler}), 
$\sigma_{\rm cr}$ is the critical stress for the column to buckle, 
$E$ is the Young modulus of the material,
$L$ is the column length, 
and the constant $c$ accounts for the end conditions of the column.
The formula indicates that, in order to obtain a column structure that can withstand a large compressive stress,
it is necessary to increase $R_g$.
Our results proved that 
imparting convex rounded sides and filleted corners 
to the cross section
do not significantly degrade the large buckling resistance
shown by strict rectangular square columns
endowed with the maximum value of $R_g$.

The section modulus, $Z$, is another key quantity for considering 
the mechanical stability of elastic columns.
$Z$ measures the strength of a column;
the higher the section modulus, the higher will be the resistance to yield under bending.
By multiplying $Z$ with the yield strength $\sigma_y$ of the constituent material of the column, 
we can calculate the upper limit bending moment, $M_y = Z \sigma_y$,
that the column can withstand without plastic deformation.
Our results show that the rounded square cross section is beneficial for securing
a large value of $M_y$ compared with the case of circular cross section,
which may be one of the reason why a kind of plant stems exhibit the rounded square cross sections.

It is also noteworthy that the second moment of area, $I$, is also important 
for evaluating the cross-sectional performance.
It represents the degree of bending stiffness of a column.
Multiplying $I$ by $E$ of the constituent material,
we obtain the bending stiffness $EI$ of the column.
Since in the present work 
the area of cross section $A$ was fixed,
the relation of $I \propto R_g^2$ holds for every systems under consideration;
in this case, the effect of $I$ on the cross-sectional performance
can be deduced easily from that result of $R_g$.

\begin{figure}[ttt]
\centering
\includegraphics[width=8.0cm]{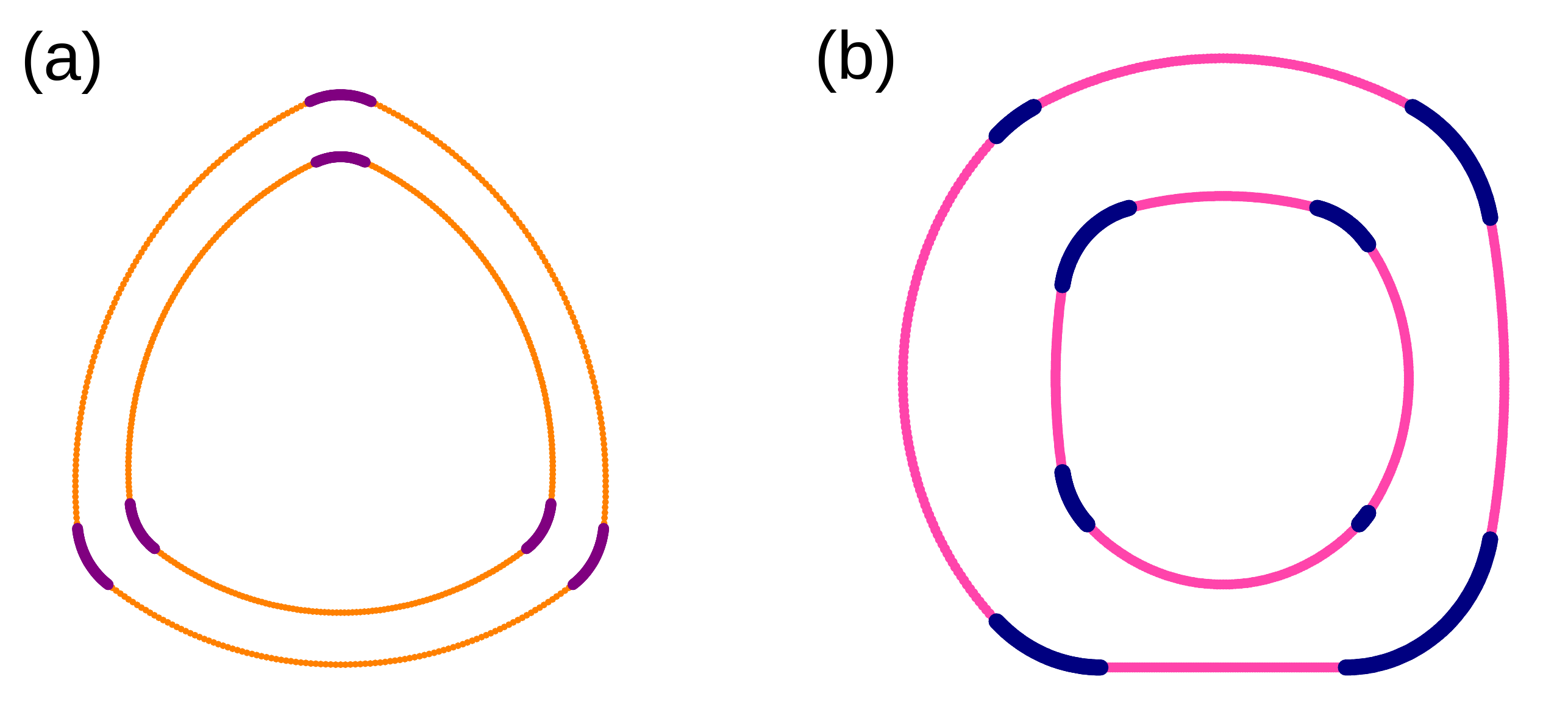}
\caption{(a) A model of triangular cross section with rounded sides and filleted corners, mimicking the cross section of a papyrus; see Fig.~\ref{fig01_plant_photo}.
(b) A model  of non-symmetric hollow square cross section.}
\label{fig06_tri_nonsym}
\end{figure}

\subsection{Versatility of the theoretical model}

As a closing remark, we mention the scope of application of our mathematical model to actual wild plants.
When you observe culms and stems of actual plants, you will find that the shape of the cross section is not uniform but gradually changes along the height direction. Therefore, the morphological quantities that characterize the cross sectional performance of the plants, such as $A, I, R_g$ and $Z$, generally change depending on the height from the ground.
Our model is effective in dealing with problems. In fact, the theoretical model developed in this study can change the cross-sectional shape continuously by properly controlling the parameter values. 
Therefore, by collecting the actual measurement data of the cross section of the plant and reproducing the shape of each cross section with different height by the model, it is possible to accurately estimate the change in cross sectional performance along the height direction. This makes it possible to analyze the three-dimensional mechanical behavior of hollow columnar plants with a polygonal cross section, such as square bamboo.

It also should be emphasized that
the model we have developed 
can be extended to triangles, pentagons, hexagons and other arbitrary polygons
in a straightforward manner.
This is realized by replacing the reference square 
depicted in Fig.~\ref{fig02_sqmodel} 
with another kind of polygon,
followed by the same drawing procedure
of long and short arcs.
Figure \ref{fig06_tri_nonsym}(a) shows
an example of a rounded triangle,
whose shape looks similar to the cross section of a papyrus ({\it Cyperus microiria})
given in Fig.~\ref{fig01_plant_photo}(d).
Furthermore, it is possible to break the discrete rotational symmetry of the model,
by setting a different value of $\tht$ at each vertex.
Figure \ref{fig06_tri_nonsym}(b) shows
such an example of rounded square with no rotational symmetry.
By extending this model in the abovementioned way, 
it becomes possible to more faithfully reproduce the cross section of 
polygonal stems, culms, and branches of actual plants
whose shape should vary considerably from sample to sample in general.
In future work, we plan to measure the cross-sectional shape of square bamboo in detail and reproduce it numerically using the present model.

\section{Conclusion}

In this article, we proposed a new mathematical model of a quadrilateral consisting of arched edges and filleted vertices. This model can accurately reproduce the cross-sectional shape of plants with hollow prismatic structures such as square bamboo, perilla and cyperus. Therefore, the model is very useful when investigating the mechanical properties of plants and when designing mechanical optimal structures that mimic the functional morphology of plants. As an application example, we showed an analytical solution of the gyration radius and the section modulus of the area of a hollow square column whose cross-sectional shape resembles that of a square bamboo. The proposed model can be applied to various polygons such as triangles, pentagons, hexagons as well as rectangles. In addition, an asymmetrical polygonal cross section can be reproduced with a slight expansion. We hope that the model will be used as an analysis tool to reproduce the morphology of actual plants with high accuracy.

\section*{Acknowledgments}
We would like to thank Dr.~T.~Fukuhara and Mr.~N.~Matsuoka 
for providing us with cross-sectional photographs of the plants.
This work was supported by JSPS KAKENHI Grant Numbers 
18H03818, 18H02244, 18KT0037, 19K03766, and 19H05359.

\appendix

\section{Preparatory materials}

Appendices A and B are devoted to derive the analytic expression of the second moment of area, $I$, of a square with rounded edges and fillet corners. Once $I$ is obtained, the gyration radius $R_g$ and the section modulus $Z$ of the same cross section are easily obtained. The following discussion presents some preparatory material for deriving the area second moment formula used in Appendix B.

\begin{figure}[bbb]
\centering
\includegraphics[width=7.5cm]{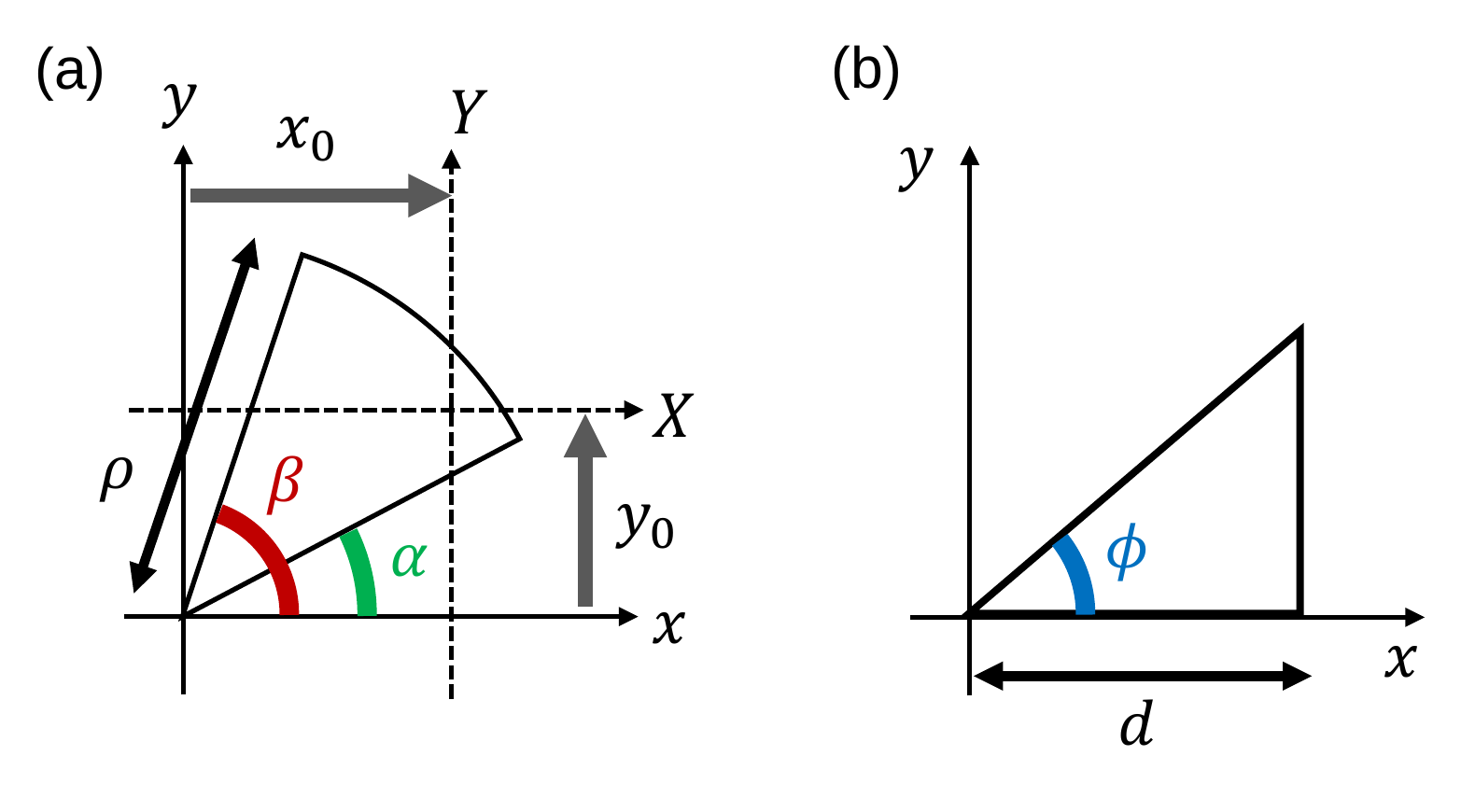}
\caption{(a) Circular sector defined by the angles $\alpha$, $\beta$, 
and the radius $\rho$.
(b) Right-angle triangle with the base $d$ and the apex $\phi$.}
\label{fig07_app_sector}
\end{figure}

For a given cross section spanned in the $x-y$ coordinate plane,
the first moment of area, $F_k$, and 
the second moment of area, $I_k$,
with respect to the $k$ axis are respectively defined by
\begin{eqnarray}
I_k &=& \iint k^2\; dx dy, \quad k=x,y, \nonumber \\
F_k &=& \iint k\; dx dy, \quad k=x,y.
\end{eqnarray}
It is easily derived from the definitions that
the second moment of area $I_X$ with respect to the $X$ axis,
which is obtained by translating the $x$ axis by $y_0$ in the vertical direction,
is represented as
\begin{equation}
I_X = I_x - y_0 F_x + y_0^2 A,
\end{equation}
where $A$ is the area of the cross section considered.
This formula will be used in Appendix B.

For later use,
two specific  kinds of geometry are considered:
the one is the circular sector depicted in Fig.~\ref{fig07_app_sector}(a),
and the other is the right triangle depicted in Fig.~\ref{fig07_app_sector}(b).
The geometry of the circular sector is uniquely determined by the radius $r$
and the two angles, $\alpha$ and $\beta$.
Hence, the second and first moments of area of the sector with respect to $k$-axis
$(k=x,y)$ are given by
\begin{eqnarray}
I_x^c(r,\alpha,\beta) &=& \frac{\rho^4}{8} 
\left[
(\beta-\alpha) - \frac{\sin 2\beta - \sin 2\alpha}{2} 
\right], \\ [6pt]
I_y^c(r,\alpha,\beta) &=& \frac{\rho^4}{8}
\left[
(\beta-\alpha) + \frac{\sin 2\beta - \sin 2\alpha}{2}
\right]
\end{eqnarray}
and
\begin{eqnarray}
F_x^c(r,\alpha,\beta) &=& \frac{\rho^3}{3} \left( -\cos \beta + \cos \alpha \right), \\ [3pt]
F_y^c(r,\alpha,\beta) &=& \frac{\rho^3}{3} \left( \sin \beta - \sin \alpha \right),
\end{eqnarray}
respectively.
The superscript $c$ indicates that the quantity is associated with
the circular sector.

Similarly, the geometry of the right triangle is uniquely determined by
the base length $d$ and the apex angle $\phi$.
Hence we have for the triangle,
\begin{eqnarray}
I_x^t(d,\phi) &=& \frac{d^4}{12} \tan \phi, \quad
I_y^t(d,\phi) = \frac{d^4}{4} \tan \phi, \\ [3pt]
F_x^t(d,\phi) &=& \frac{d^3}{6} \tan \phi, \quad
F_x^t(d,\phi) = \frac{d^3}{3} \tan \phi.
\end{eqnarray}
The superscript $t$ indicates that the quantity is associated with
the triangle.

\section{Formulae of the second moment of area}

\begin{figure*}[ttt]
\centering
\includegraphics[width=11.5cm]{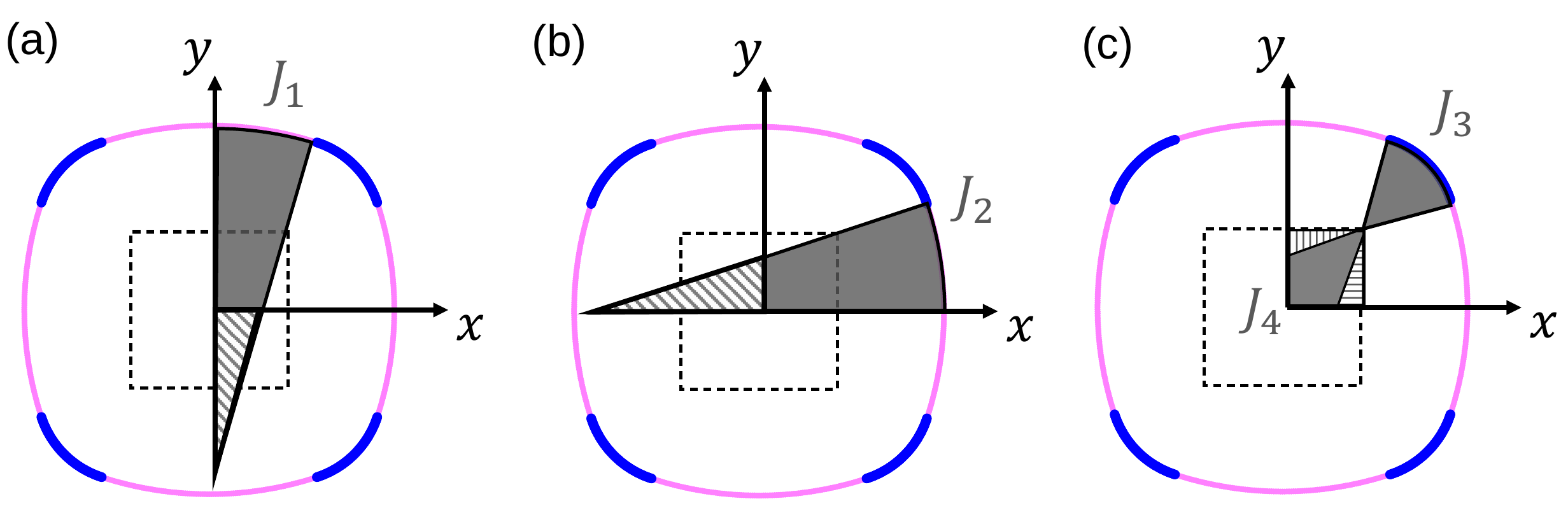}
\caption{Diagram of the four domains in the first quadrant ($x>0$ and $y>0$) 
of the rounded square. Each domain is marked by a gray area.
$J_i$ $(i=1,2,3,4)$ represents the second moment of area of the $i$-th domain
with respect to the $x$-axis.
(a) Domain 1; the gray area obtained by excluding the shaded right triangle
from the vertically elongated sector.
(b) Domain 2; the gray area obtained by excluding the triangle from the long sector.
(c) Domains 3; the gray small sector at the upper right,
and Domain 4; the gray kite-shaped square
obtained by excluding the two shaded triangles from the right square.}
\label{fig08_app_domain}
\end{figure*}

Our immediate goal is to formulate 
the second moment of area of the rounded-corner squared cross section,
defined by the previous section,
with respect to the $x$-axis.
To the aim,
we decompose the first quadrant part of the cross section into the four domains
as illustrated by Fig.~\ref{fig08_app_domain}.
For the four domains, the second moments of area with respect to the $x$-axis
are written by $J_i$ $(i=1,2,3,4)$;
then, the second moment of the whole square, $J_{\rm all}$,
is given by
\begin{equation}
J_{\rm all} = 4 \left( J_1 + J_2 + J_3 - J_4 \right).
\end{equation}
In the following, we will show how to calculate the four components
$J_i$ $(i = 1,2,3,4)$ in order.

For the domain 1, the second moment of area with respect to the $x$-axis is given by 
\begin{equation}
J_1 = J_{1X}^c - J_{1X}^t,
\end{equation}
where $J_{1X}^c$ is the second moment of the circular sector
which is elongated in the $x$-direction, as depicted in Fig.~\ref{fig08_app_domain}(a),
and $J_{1X}^t$ is that of the right triangle (shaded region) elongated in the $x$-direction, too.
The explicit forms of them read
\begin{eqnarray}
J_{1X}^c &=& I_x^c(\rho_1,\alpha_1,\beta_1) - 2 d F_x^c(\rho_1,\alpha_1,\beta_1) + d_1^2 A_1^c, \nonumber \\
J_{1X}^t &=& I_y^t(d_1,\phi_1)- 2 d F_y^t(d_1,\phi_1) + d_1^2 A_1^t.
\label{eq_app_045}
\end{eqnarray}
Here, the subscript 1 attached to the arguments
indicate that their value equal to those listed in the row of ``Domain 1"
in Table \ref{table_01}.
$A_1^c$ and $A_1^t$ are the area of the circular sector and the triangle,
respectively,
given by $A_1^c = (\phi_1/2) \rho_1^2$
and $A_1^t = (d_1^2/2) \tan [ (\pi/2) - \tht]$.
The only exception is the case in which $\tht=\pi/2$.
In this case, both $\rho_1$ and $d_1$ diverges so that Eq.~(\ref{eq_app_045}) does not work properly.
Instead, we should use the different expression of
$J_1 = (\ell/6) [(\ell/2) + h]^3$,
only when $\tht=\pi/2$.

\renewcommand{\arraystretch}{1.6}
\begin{table}[bbb]
  \centering
  \caption{Parameter settings for the four domains.}
  \begin{tabular}{|c|c|c|c||c|c|} \hline
    Parameters & $\rho$ & $\alpha$ & $\beta$ & $d$ & $\phi$ \\ \hline \hline
    Domain 1   & $\displaystyle{\frac{\ell}{2 \cos \tht} + h}$ & $\tht$ & $\displaystyle{\frac{\pi}{2}}$ & $\displaystyle{\frac{\ell}{2} (\tan \tht-1)}$ & $\displaystyle{\frac{\pi}{2} - \tht}$ \\ [5pt]
    Domain 2   & $\displaystyle{\frac{\ell}{2 \cos \tht} + h}$ & $0$ & $\displaystyle{\frac{\pi}{2} -\tht}$ & $\displaystyle{\frac{\ell}{2} \left( \tan \tht - 1 \right)}$ & $\displaystyle{\frac{\pi}{2} - \tht}$  \\ [5pt]
    Domain 3   & $h$ & $\displaystyle{\frac{\pi}{2} - \tht}$ & $\tht$ & ** & **  \\ [5pt]
    Domain 4   & ** & ** & ** & $\displaystyle{\frac{\ell}{2}}$ & $\displaystyle{\frac{\pi}{2} - \tht}$  \\ \hline
  \end{tabular}
\label{table_01}
\end{table}
\renewcommand{\arraystretch}{1.0}

For the domain 2, the second moment of area, $J_2$, is given by
\begin{equation}
J_2 = I_x^c(\rho_2,\alpha_2,\beta_2)-I_x^t(d_2,\phi_2),
\end{equation}
for $\pi/4 \le \tht < \pi/2$.
Only when $\tht=\pi/2$, it is replaced by 
$J_2 = (\ell^3/24) [(\ell/2)+ h]$.

For the domain 3, we have
\begin{equation}
J_3  =
I_x^c(\rho_3,\alpha_3,\beta_3) + \ell F_x^c(\rho_3,\alpha_3,\beta_3)
+ \left( \frac{\ell}{2} \right)^2 A_3^c,
\end{equation}
where $A_3^c = (\rho_3^2/2) (\beta_3-\alpha_3)$.
This expression applies regardless of the value of $\tht$,
as long as $\pi/4 \le \tht \le \pi/2$.

For the domain 4, we have
\begin{equation}
J_4 = \frac{\ell^4}{48} - \left( J_{\rm 4V}^t + J_{\rm 4H}^t \right).
\end{equation}
Here, $J_{\rm 4V}^t$ is the second moment of area of the vertically elongated triangle,
and $J_{\rm 4H}^t$ is that of the horizontally elongated triangle,
both of which are shaded in Fig.~\ref{fig08_app_domain}(c).
They are explicitly written by
\begin{eqnarray}
J_{\rm 4V}^t &=& I_y^t(d_4,\phi_4) - \ell F_y^t(d_4,\phi_4) 
+ \left( \frac{\ell}{2} \right)^2 A_4^t, \\
J_{\rm 4H}^t &=& I_x^t(d_4,\phi_4) - \ell F_x^t(d_4,\phi_4) 
+ \left( \frac{\ell}{2} \right)^2 A_4^t,
\end{eqnarray}
where $A_4^t = (\ell^2/8) \tan \phi_4$ is the area of the shaded triangle.

\section{Isotropy of $I$ for the four-fold symmetric cross-section}

It proves that the value of $I$ is uniquely determined only by the shape of the cross section, regardless of axial choice, when the cross section is four-fold symmetric.

Suppose that the coordinate axes in the original $x$-$y$ plane
are mutually rotated by the angle $\varphi$.
The new coordinates are defined by
\begin{eqnarray}
u &=& x \cos\varphi + y \sin \varphi, \nonumber \\
v &=& -x \sin\varphi + y \cos \varphi.
\end{eqnarray}
The second moments of area with respect to the rotated axes read
\begin{eqnarray}
I_u &=& I_x \cos^2 \varphi + I_y \sin^2 \varphi - I_{xy} \sin 2\varphi, 
\label{eq_app_085} \\
I_v &=& I_x \sin^2 \varphi + I_y \cos^2 \varphi + I_{xy} \sin 2\varphi,
\label{eq_app_087}
\end{eqnarray}
where $I_{xy}$ was defined by
\begin{equation}
I_{xy} = \iint xy\; dx dy.
\end{equation}
Remind that when the cross section to be considered
is endowed with a four-fold symmetry in the $x$-$y$ plane,
we have 
\begin{equation}
I_x = I_y \;\; {\rm and} \;\; I_{xy}=0.
\end{equation}
Substituting them into Eqs.~(\ref{eq_app_085}) and (\ref{eq_app_087}),
we achieve the conclusion that
\begin{equation}
I_x = I_u = I_v = I_y,
\end{equation}
which hold true for arbitrary $\varphi$.
This result indicates that the second moment of the four-fold symmetric cross-sectional area does not depend on the rotation angle of the axis and is uniquely determined by the cross-sectional shape.

\bibliographystyle{apsrev4-1}
\bibliography{Gyration_Shima_arXiv}

\end{document}